\documentclass[twocolumn,a4paper,floatfix]{revtex4}

\usepackage{graphicx}
\usepackage{multirow}
\usepackage{amsmath}
\usepackage{color}
\usepackage{epsfig}
\usepackage{epstopdf}

\begin{document} 
\title{Study of the ST2 model of water close to the liquid-liquid critical point}

\author{Francesco Sciortino}
\affiliation{Dipartimento di Fisica,
Universit\`a di Roma {\em La Sapienza},
Piazzale A. Moro 5, 00185 Roma, Italy}
\author{Ivan Saika-Voivod}
\affiliation{Department of Physics and Physical Oceanography,
Memorial University of Newfoundland, St. John's, Newfoundland and Labrador, A1B 3X7, Canada}
\author{Peter H. Poole}
\affiliation{Department of Physics, St. Francis Xavier University, 
Antigonish, Nova Scotia B2G 2W5, Canada}

\begin{abstract}
We perform successive umbrella sampling grand canonical Monte Carlo 
computer simulations of the original ST2 model of water in the vicinity of the proposed liquid-liquid critical point, at temperatures above and below the critical temperature.
Our results support the previous work of Y. Liu, A.Z. Panagiotopoulos and P.G. Debenedetti [J. Chem. Phys. {\bf 131}, 104508 (2009)], who provided evidence for the existence and location of the critical point for ST2 using the Ewald method to evaluate the long-range forces. Our results therefore demonstrate the robustness of the evidence for critical behavior with respect to the treatment of the electrostatic interactions.  
In addition, we verify that the liquid is equilibrated at all densities on the Monte Carlo time scale of our simulations, and also that there is no indication of crystal formation during our runs.  
These findings demonstrate that the processes of liquid-state relaxation and crystal nucleation are well separated in time.  Therefore, the bimodal shape of the density of states, and hence the critical point itself, is a purely liquid-state phenomenon that is distinct from the crystal-liquid transition.  
%In addition, we find no contribution of crystallization to the bimodal shape of the density of states, and that the liquid is readily equilibrated at all densities on the time scale of our simulations, thus demonstrating that the critical point is a purely liquid state phenomenon and is thermodynamically distinct from the crystal-liquid transition.  
%We find  that the coexisting high density liquid becomes denser on cooling, while for the low density liquid phase, the density stays relatively constant, reflecting the optimal density required for the formation of a tetrahedral network.
\end{abstract}

\maketitle

%\keywords{statistical physics | colloids | simulations}

\section{Introduction}
In 1992, a numerical investigation of the equation of state (EOS) of the 
ST2 model~\cite{ST2} in the supercooled 
region suggested the  possibility of a liquid-liquid (LL) critical point in water~\cite{pses92}. 
This initial study has subsequently generated a large amount of numerical and 
experimental work~\cite{hes17,hes26, abascal2010, Harrington:1997p4374,soper2000,hes29,lanave03,mou2005,Loerting:2006p4014,loerting}. 
In addition to the conceptual novelty of a one-component system with more than one liquid phase,  the existence of the associated LL critical point can also rationalize many of the thermodynamic anomalies which characterize liquid water (e.g. the density maximum and compressibility minimum), and which become more pronounced in the supercooled regime.
Furthermore, the existence of two distinct liquid phases of supercooled water can explain the
polyamorphism which characterizes the glassy phase~\cite{mishima94,mishima2000,loerting}. Indeed, simulations suggest that the low density amorphous solid form of water is similar to the structure of the low density liquid (LDL) phase, while the relaxed very-high density amorphous solid is related to the high density liquid (HDL)\cite{nicolasPRE2005}.

Evidence in support of a liquid-liquid critical point in water, and in other liquids with tetrahedral structure, has increased over time. A number of classical models for water, including the recently developed and optimized TIP4P/2005~\cite{vega},  exhibit a van~der~Waals inflection in their EOS at low temperature $T$ that is evidence of 
phase coexistence between two liquid states~\cite{paschek2005,gallo2010,abascal2010}.   
The occurrence of a LL transition has also been proposed for silica~\cite{saikapre2001}, and
more recently, evidence for a LL critical point and its associated thermodynamic anomalies have been presented for the Stillinger-Weber model of silicon~\cite{sastrynature}.  

Indeed, it is notable that the most compelling evidence for LL critical points has been
generated {\it in silico}~\cite{nandviews}. In almost all cases, LL critical points are predicted to occur in deeply supercooled liquids, where crystallization (in experiments) has so far prevented direct observation of such phenomenon in bulk systems.  Compared to experiments, LL phase transitions are more readily observed in numerical studies because heterogeneous 
nucleation is not a factor, and the small system size (usually less than one thousand molecules)
decreases the probability of observing the appearance of a critical crystal nucleus in the simulation box on the time scale of typical simulations.   Computer simulations have thus allowed the study of the liquid EOS under deeply supercooled conditions, on time scales longer than the structural relaxation time of the liquid but smaller than the homogenous nucleation time. Under these conditions, equilibrium within the metastable basin of the liquid free energy surface can be achieved without interference from crystal nucleation processes.
%~\cite{Poole:2005p2770}. 

Nonetheless, evaluations of the EOS via simulations in the canonical ensemble, or at constant pressure, do not provide a way to accurately estimate the
location of the LL critical point found in water models, or to determine its universality class.
Only recently, in 2009, Liu, et al.~\cite{lpd09} reported the first numerical investigation 
of ST2 water in the LL critical region, performing simulations in the grand canonical ensemble for different values of $T$ and of the chemical potential $\mu$, and
implementing Ewald sums to account for the long-range contributions to the electrostatic interactions. In this important contribution, the authors  provided for the first time evidence of a density of states that is a bimodal function of the density $\rho$, a necessary feature of a LL critical point. Importantly, the authors also showed that the fluctuations of the order parameter
(a combination of density and energy) are consistent with the expected shape
for a critical system in the Ising universality class.

More recently, Limmer and Chandler~\cite{limmer3} have questioned the interpretation of all previously published simulations based on the ST2 potential, arguing that ``that behaviors others have attributed to a liquid-liquid transition in water and related systems are in fact reflections of transitions between liquid and crystal.''
In the case of the recent calculations of Liu, et al.~\cite{lpd09}, Ref.~\cite{limmer3} proposes that  ``the Liu et al. result is a non-equilibrium phenomenon, where a long molecular dynamics run initiated from their low-density amorphous phase and run at constant $T$ and $P$ will eventually equilibrate in either the low density crystal or (more likely) in the higher density metastable liquid.''
It is thus of paramount importance to independently check the findings of Liu, et al.~\cite{lpd09}, and at the same time, test whether or not the LDL phase is truly a disordered liquid phase characterized by a well-defined metastable equilibrium that is distinct from the crystal phase.

In this article, we conduct these tests by carrying out an independent
evaluation of the density of states based on the successive umbrella sampling technique~\cite{sus}.  We implement the original ST2 model, with the reaction field correction for the
long range electrostatic forces, rather than Ewald sums, to be able to strictly compare our results with previously published data for ST2~\cite{Poole:2005p2770}, as well as to test if the LL transition is robust and independent of the treatment of the long range interactions.  As we show below, we find that our results are entirely consistent with those of Liu, et al.~\cite{lpd09}, as well as with earlier simulation data.  We further find that there is no contribution to the density of states due to crystal formation, confirming the distinct existence of both the HDL and LDL phases for $T$ less than $T_c$, the temperature of the LL critical point.

\section{model and simulation methods}

We study the original ST2 potential as defined by Rahman and Stillinger~\cite{ST2}, with reaction field corrections to approximate the long-range contributions to the electrostatic interactions.  In the ST2 potential, water is modeled as a rigid body with an oxygen atom at the center and four charges, two positive and two negative, located at the vertices of a tetrahedron.  The distances from the oxygen to the positive and negative charges are 0.1 and 0.08~nm, respectively.  The oxygen-oxygen interaction is modeled using a Lennard-Jones potential with $\sigma_{LJ}=0.31$~nm and 
$\epsilon_{LJ}=0.316 94$ kJ/mol.  We truncate this Lennard-Jones interaction at $2.5 \sigma_{LJ}$, 
accounting for the residual interactions through standard long range corrections, i.e. assuming 
the radial distribution function can be approximated by unity beyond the cutoff.  The charge-charge interactions are smoothly switched off both at small and large distances via a tapering function, as in the original version of the model~\cite{ST2}. 

Our grand canonical Monte Carlo (MC) algorithm
is based on roto-translational moves, insertions, and deletions, each attempted with ratios
2:1:1.
Our simulation box is cubic with sides of length $2$~nm.
The  displacement move is accomplished by a random translation in each direction of
up to  $\pm 0.01$~nm and a 
random rotation of up to $\pm 0.2$~rad around a random direction, resulting
in an acceptance ration of about 50$\%$.  Insertion and deletion moves have 
a much smaller acceptance ratio, of the order of $10^{-5}$.  The simulations have thus
been performed for more than $10^{10}$  attempted insertion/deletion moves.
To determine the dependence of the density of states on $T$ we have investigated
four distinct temperatures, $T=260$, $250$, $245$ and $240$~K. 
Previous numerical estimates based on the EOS indicate $T_c = 247 \pm 3$ K~\cite{Poole:2005p2770,megan}.

To study the phase behavior of the system we implement successive umbrella 
sampling (SUS) MC simulations~\cite{sus}, from which we evaluate the 
probability density $P(\rho)$ for the values of $\rho$ sampled by the equilibrium system at fixed $T$, $\mu$, and volume $V$, and in which the number of molecules $N$ in the system fluctuates.
In the SUS method, the pertinent range of $\rho$ to be investigated, written in terms of the lower and upper number of molecules (respectively $N_l$ and $N_u$), is divided into  many 
small overlapping windows of size $\Delta N$. For each window, a separate grand canonical 
MC simulation monitors how often a state of $N$ particles is visited.
Moreover, the simulations are constrained using appropriate boundary conditions on $N$,
such that deletions or insertions that would cause $N$ to vary outside the range assigned to that window are rejected~\cite{binder}. The density histograms for each window can then be combined to obtain the full $P(\rho)$ curve, by imposing the equality of the probability at the overlapping 
boundary.   In our study $N_l=200$ (corresponding to a minimum density
$\rho=0.75$ g/cm$^3$) and $N_u=327$ (corresponding to $\rho=1.22$ g/cm$^3$).
We have chosen $\Delta N=2$, i.e. $N$ is only permitted to take on one of two adjacent integer values within each window. 

The SUS method has a number of advantages.
The use of narrow windows in $N$ allows an effective sampling of the microstates without the use of
biasing functions. Since the windows are independent, all the simulations
can be run in parallel, with a gain in throughput that scales linearly with the 
number of processors employed. In our case, approximately 130 processors are used 
for the calculations at each $T$, one for each window. At the lowest $T$, more than two months of simulation time for each window is required for good sampling.
Once we obtain $P(\rho)$ at fixed $T$ and $\mu$, histogram reweighting techniques~\cite{reweight} 
can be applied to obtain $P(\rho)$ at different values of $\mu$.  Keeping track of the coupled density-energy histogram during the SUS simulations 
also allows us to estimate $P(\rho)$ at different $T$ via temperature reweighting. 
%Histogram reweighting techniques are  fundamental to locating both the critical point and the coexistence points.
Finally, each window provides accurate information on a specific density, allowing us to  
compare the results with EOS data from previous simulations in the canonical ensemble.

To facilitate comparison of our data with future studies, in the following we will report 
the activity as  $z^* \equiv  \Lambda_x \Lambda_y \Lambda_z  \exp(\beta \mu)/\lambda^3$ 
(in units of $nm^{-3}$), where  $\mu$ is the chemical potential, $\beta$ is $(k_BT)^{-1}$, $k_B$ is the Boltzmann constant, $\Lambda_\alpha \equiv \sqrt{(2\pi I_{\alpha} k_BT)}/h$,
$I_{\alpha}$ is the principal moment of inertia in the direction $\alpha$, $h$ is the Planck
constant and   $\lambda$ is the De Broglie wavelength.  $z^*$ is the quantity that enters in the MC acceptance probability in the insertion-deletion moves. 

%confirming that the SUS simulations reproduce previous simulations results.
%testing how the SUS results converge to those obtained from previous simulations. 

%The coexistence points at fixed $T$ are obtained by reweighting the densities histogram until the two peaks (for the low- and high-density phases) have the same area and their average density provide precise estimates of the coexistence densities.

\section{Comparison with previously published data}

\begin{figure} 
\hbox to \hsize{\epsfxsize=1.0\hsize\epsfbox{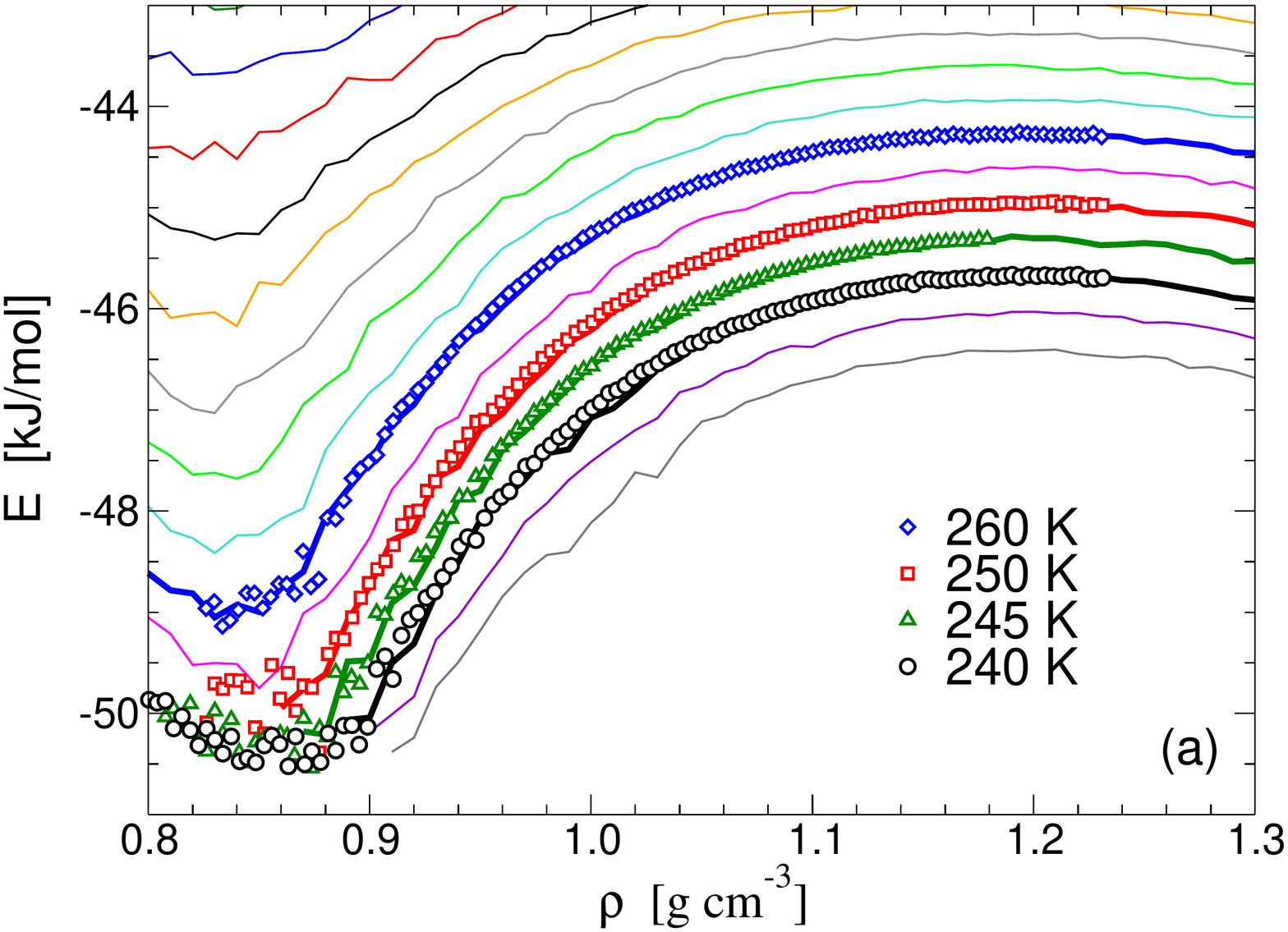}}
\hbox to \hsize{\epsfxsize=1.0\hsize\epsfbox{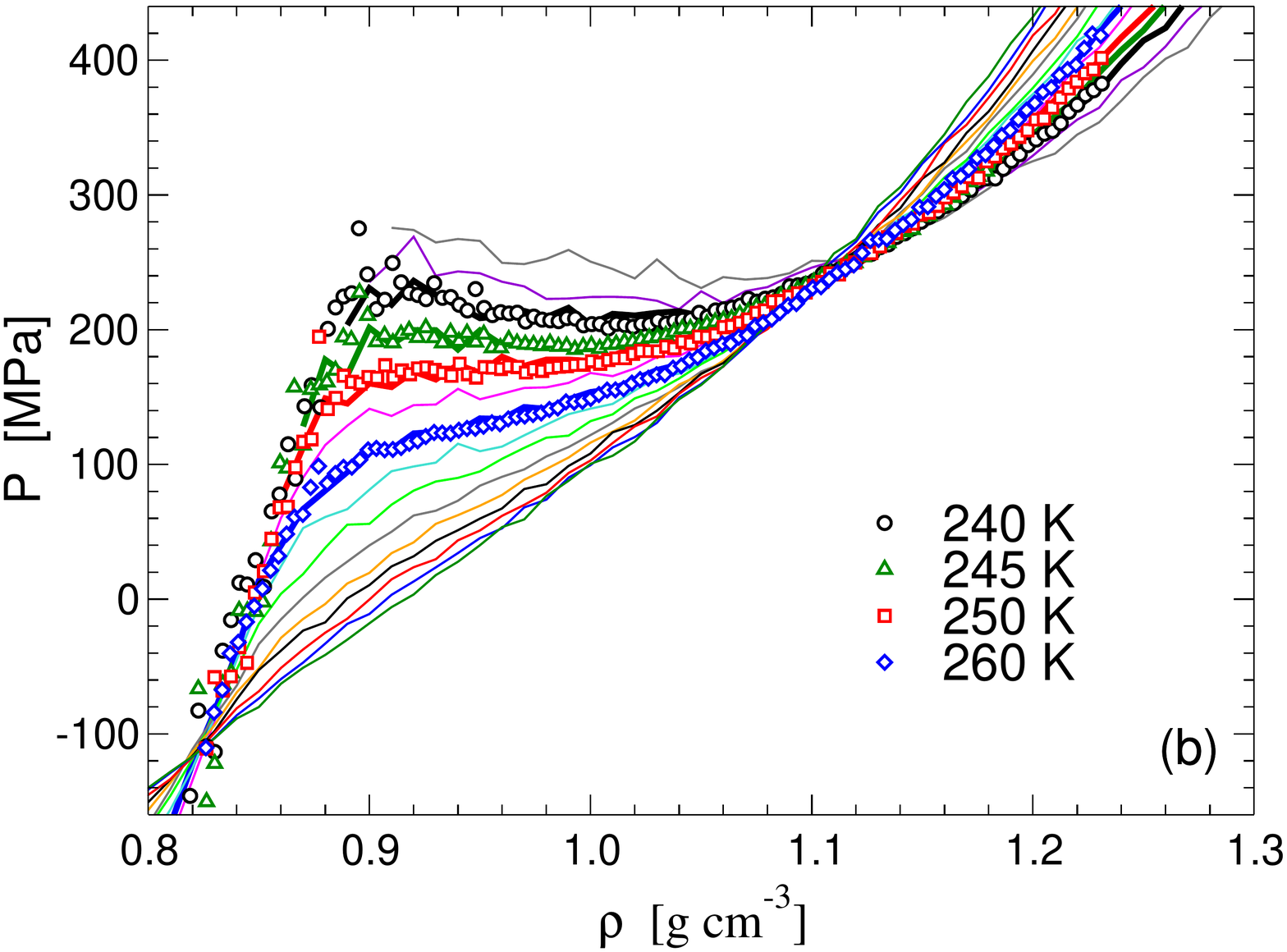}}
\caption{Consistency of results with previously published MD simulation data~\cite{Poole:2005p2770,preprint}.
Shown are isotherms of the (a) potential energy and (b) pressure.  Open symbols indicate data
obtained from the individual SUS windows.  Curves show previous MD results, starting from $T=230$~K and spaced every 5~K.  Curves are thicker for $T$ matching the SUS simulations.
%Panel (a) shows potential energy as a function of density along isotherms.
%Panel (b) shows the pressure.
}
\label{fig:comparison}
\end{figure}

Before discussing the behavior of $P(\rho)$, we compare our
results for the liquid EOS as obtained from our SUS simulations, with the best available published data. In particular, we focus on the potential energy $E$ and the pressure $P$.  The pressure is evaluated from the virial using configurations sampled at each density.  Fig.~\ref{fig:comparison} shows results from Ref.~\cite{Poole:2005p2770}, obtained from molecular dynamics (MD) simulations, compared with our MC grand-canonical SUS results. At all $T$ and $\rho$, we find excellent agreement between these two completely independent numerical methods.  

It is worth noting the prominent minimum observed in $E$
near $\rho=0.83$ g/cm$^3$, the so-called optimal network density~\cite{pwmnoi,preprint}.
At this density the system has the possibility to be able to satisfy all possible bonds, reaching 
at low $T$ the ideal random tetrahedral network state. At lower densities, gas-liquid
phase separation intervenes, while at larger densities closer packing
prevents the system from satisfying the angular and distance constraints 
required to form linear hydrogen bonds between all pairs of molecules.   

\section{Liquid-state equilibrium}

In order to establish that the LL phase transition is a genuine liquid-state phenomenon, we must confirm (i) that all our simulations are carried out over a time scale that is much longer than the structural relaxation time of the liquid; and (ii) that the time scale for crystal nucleation is much longer than the time scale for liquid-state relaxation.  The separation of these two time scales provides the ``window'' within which the equilibrium behavior of a supercooled liquid can be defined and quantified.

The ability of modern computer simulations to establish liquid-state equilibrium in simulations near a LL critical point is well documented~\cite{paschek2005,gallo2010,preprint,Poole:2005p2770,sastrynature,vega}.  In particular, Ref.~\cite{preprint} presents the dynamical behavior of the same ST2 model as is studied here, as determined from MD simulations.  In Fig.~2(a) of Ref.~\cite{preprint}, it is shown that the self diffusion constant $D$ for oxygen atoms is greater than $10^{-8}$~cm$^2$/s at $T=240$~K for all densities from $0.87$ to $1.2$~g/cm$^3$.  This density range spans the same range within which we find bimodal behavior for $P(\rho)$ at $T=240$~K (see below).  For the ST2 system, $D>10^{-8}$~cm$^2$/s corresponds to a range of $\alpha$-relaxation times $\tau_\alpha<20$~ns~\cite{Becker:2006p15}.  Time scales well in excess of $20$~ns are readily accessible in current MD simulations, especially for a small system of a few hundred molecules, as is studied here.  

\begin{figure} 
\hbox to \hsize{\epsfxsize=1\hsize\epsfbox{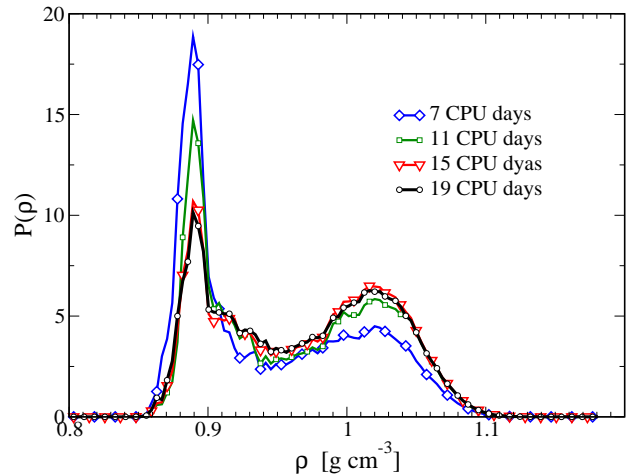}}
\caption{Example of the convergence of the density of states during SUS MC simulations.
Shown is $P(\rho)$ for $T=245$~K and $z^*=1.46\times 10^{-4} $ nm$^{-3}$
at various times over 19 CPU-days of simulation time.}
\label{fig:eq}
\end{figure}

Correspondingly, MC simulations of the kind reported here can also easily be run for the number of MC steps required to access equilibrium liquid properties.  To demonstrate this, we show in Fig.~\ref{fig:eq} the $P(\rho)$ histograms generated in our simulations as a function of the computing time invested, for $T=245$~K at a fixed value of the activity $z^*$.  We find that $P(\rho)$ converges to well-defined values for all densities considered here.
In addition we also check that the energy autocorrelation function decays to zero, in all
SUS windows, in a number of steps smaller than the simulation length. 

We also note that Ref.~\cite{preprint} demonstrates that the dynamical properties of the liquid state of ST2 in the low density region near the optimal network density ($0.83$~g/cm$^3$) have a very simple and well-defined dependence on $T$, including in the region where $T$ approaches $T_c$ and the LDL emerges as a distinct phase.

%We also note that previous investigations of the dynamical properties of ST2 (both translational and rotational) carried out using MD simulation show that equilibration, as monitored by the diffusion of molecules over distances larger than their diameter, has been achieved in the vicinity of the LL critical point~\cite{preprint}.  Equilibrium properties of other water models near the LL critical point also seem to be accessible on simulation timescales~\cite{paschek2005,gallo2010}. 

%We also note that previous investigations of the dynamical properties (both translational and rotational) carried out using molecular dynamics simulation~\cite{preprint,Poole:2005p2770,others} show that at $T=240$ equilibration (monitored by the decay of the density fluctuations or the diffusion of molecules over distances larger than their diameter) has been achieved\cite{preprint}.
 
\section{Analysis of the structure of the liquid in relation to the crystal}

Having shown in the previous section that liquid equilibrium is established in our simulations, we next test if crystal nucleation is occurring on the MC time scale of our runs.  This is particularly relevant in view of the conclusions presented in Ref.~\cite{limmer3}, in which it is argued that the LDL phase does not exist and that the only well-defined state of the system at low density is the crystal.  We therefore must quantify the degree of crystalline order in our system at low $T$, especially in the LDL region, where crystallization might occur on a shorter time scale due to the similarity between ice and the liquid in terms of density and local structure.  

\begin{figure}
\hbox to \hsize{\epsfxsize=1.0\hsize\epsfbox{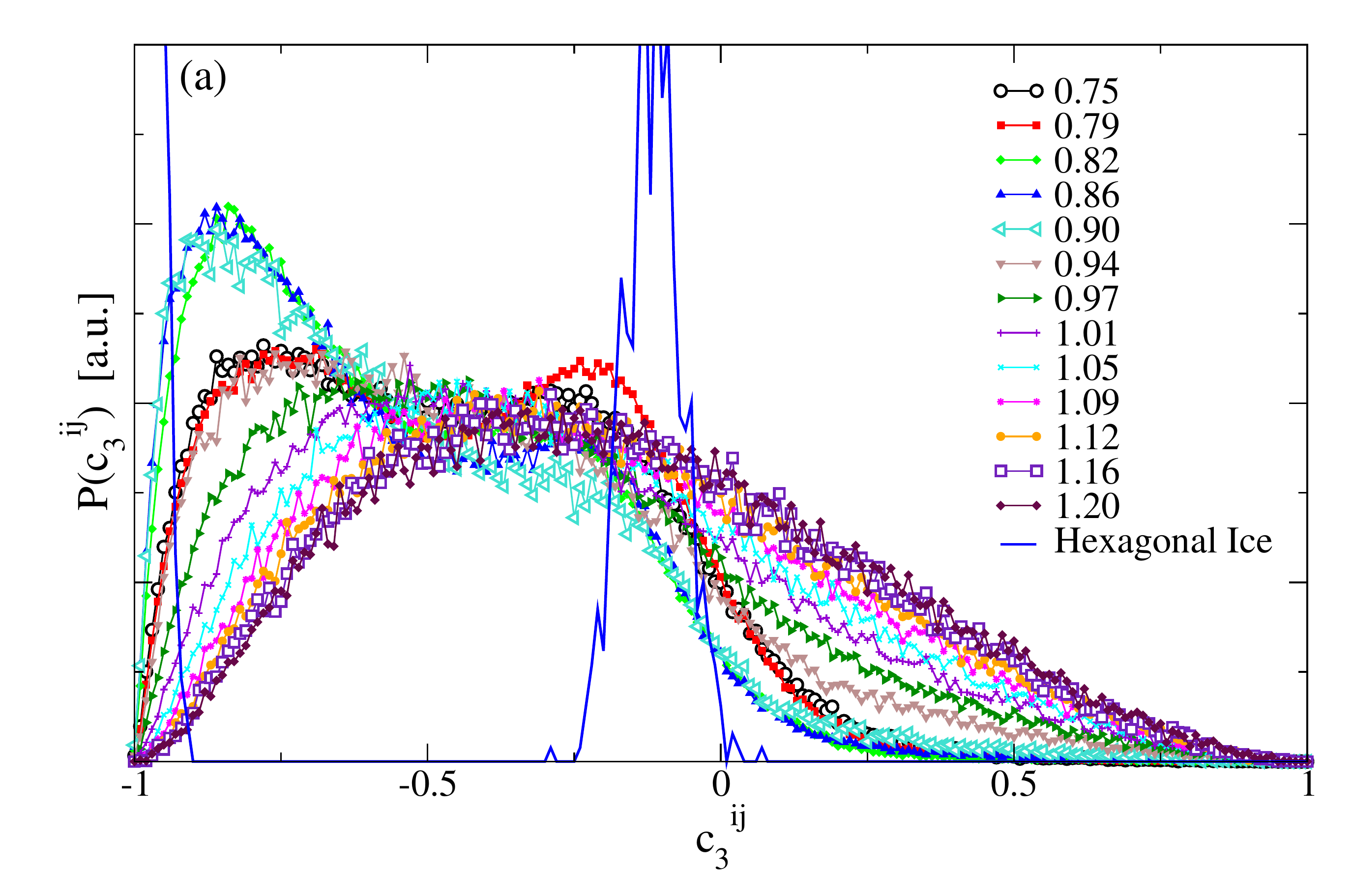}}
\hbox to \hsize{\epsfxsize=1.0\hsize\epsfbox{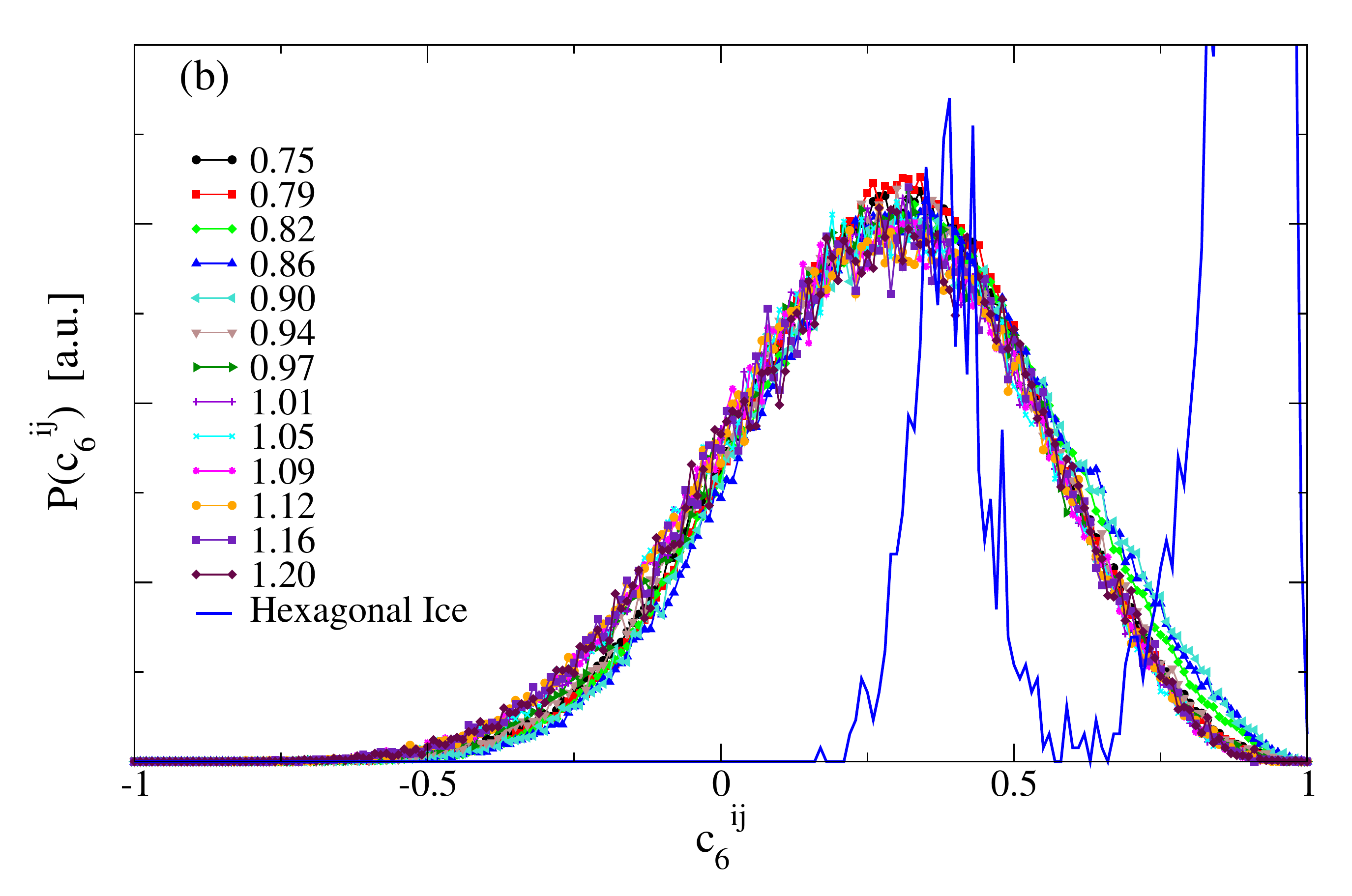}}
\hbox to \hsize{\epsfxsize=1.0\hsize\epsfbox{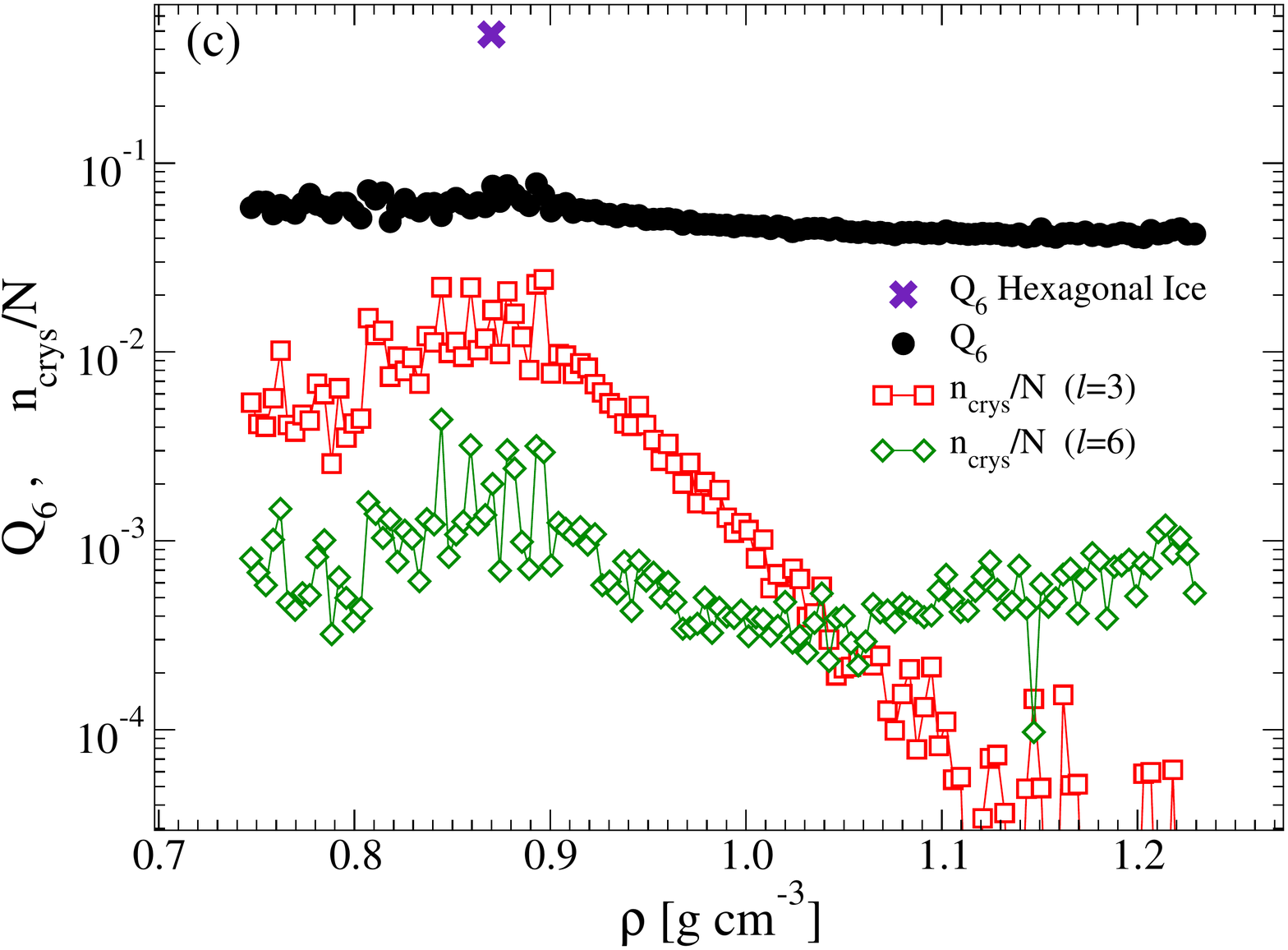}}
\caption{Characterization of crystallinity in our system at $T=240$~K.
Panel (a) shows the distribution of bond correlation dot products $P(c_3^{ij})$ obtained using order $l=3$ spherical harmonics for different SUS windows spanning our density range.  Legend labels indicate $\rho$ in each window.  Also shown is the distribution for hexagonal ice at $\rho=0.87$~g~cm$^{-3}$.  Panel (b) is the equivalent of (a), only using  $l=6$.  Panel (c) shows $Q_6$ as a function of $\rho$ for our simulations, as well as for hexagonal ice.  Also shown is the average fraction of crystalline particles in our system, using criteria based on both $l=3$ and $l=6$.
}
\label{fig:cij}
\end{figure}

The degree of crystalline order can be quantified using
the Steinhardt bond order parameters~\cite{Steinhardt} 
based on spherical harmonics of order $l=3$ and $l=6$,
which are particularly suited for discriminating disordered fluid configurations from the open structure of hexagonal (as well as cubic) ice.
For each particle we define the complex vector,
\begin{equation}
q_{lm}(i)=\frac{1}{N_b(i)}\sum_{j=1}^{N_b(i)}Y_{lm}(\hat{r}_{ij}),
\end{equation}
where the sum is over the $N_b(i)$ neighbors of particle $i$.  $Y_{lm}$ is a spherical harmonic of order $l$ and $m$, and $\hat{r}_{ij}$ is a unit vector pointing from the oxygen atom on molecule $i$ to that on molecule $j$.  In the case of
$l=3$ two molecules are considered neighbors if their  oxygen-oxygen distance is smaller than $0.34$~nm, the position of the first minimum of the radial distribution function.
In the case of $l=6$, to be consistent with Ref.~\cite{limmer3}, we assume  $N_b(i)=4$
and neighbors are defined as the four closest particles.
 The dot product,
\begin{equation}
c_l^{ij} = \sum_{m=-l}^l \hat{q}_{lm}(i) \hat{q}^*_{lm}(j),
\end{equation}
where,
\begin{equation}
\hat{q}_{lm}(i) = q_{lm}(i)/\left(\sum_{m=-l}^l \left| q_{lm}(i)\right|^2  \right)^{1/2}
\end{equation}
and
$ \hat{q}^*_{lm}(i) $ is its complex conjugate, determines the degree of orientational
correlation between neighboring particles $i$ and $j$.   Fig.~\ref{fig:cij} shows the distributions 
of $c_l^{ij}$ for $l=3$ and $l=6$, for several densities at  $T=240$ K. 
The distributions have a single broad peak at higher $\rho$, and become more bimodal at lower $\rho$ with a 
peak forming near $c_3^{ij}=-1$.  However, the distribution goes to zero at  $c_3^{ij}=-1$.
The bimodal shape of the  $c_3^{ij}$ distribution at low $\rho$ is characteristic of liquids with well 
formed tetrahedral networks~\cite{ivanJCPpreprint}.
The figure also shows the same distribution evaluated in a hexagonal ice configuration at the same $T$.  
The crystal is characterized by a large peak centered at $c_3^{ij} \approx -1$ and a smaller peak 
%(with weight $1/4$) 
located approximately at $c_3^{ij} \approx -0.1$, with $1/3$ the area of the large peak~\cite{Romano}.  In cubic ice, only the
peak at $c_3^{ij} \approx -1$ is present.   The distribution of  $c_6^{ij}$ is rather 
density independent, but again very different from the crystalline one. 

Ref.~\cite{limmer3} specifically investigated the global metric,
\begin{equation}
Q_{l,m}=\sum_{i=1}^N q_{l,m}^i
\end{equation}
and the related $m$-independent rotational invariant,
\begin{equation}
Q_l= \frac{1}{N}  \left(\sum_{m=-l}^{l} Q_{l,m} Q_{l,m}^*\right)^{1/2}.
\end{equation}
Fig.~\ref{fig:cij}(c) shows $Q_6$ at $T=240$ K evaluated from the different windows
of densities as well as for hexagonal ice.   Again, the configurations sampled in our SUS simulations do not show any sign of crystalline order.
The value of $Q_6$  in the liquid state, expected to be zero in the thermodynamic limit,  is always small ($Q_6 \approx 0.06$) and is one order of magnitude smaller than the crystal value.

\begin{figure}
\includegraphics[width=7.5cm]{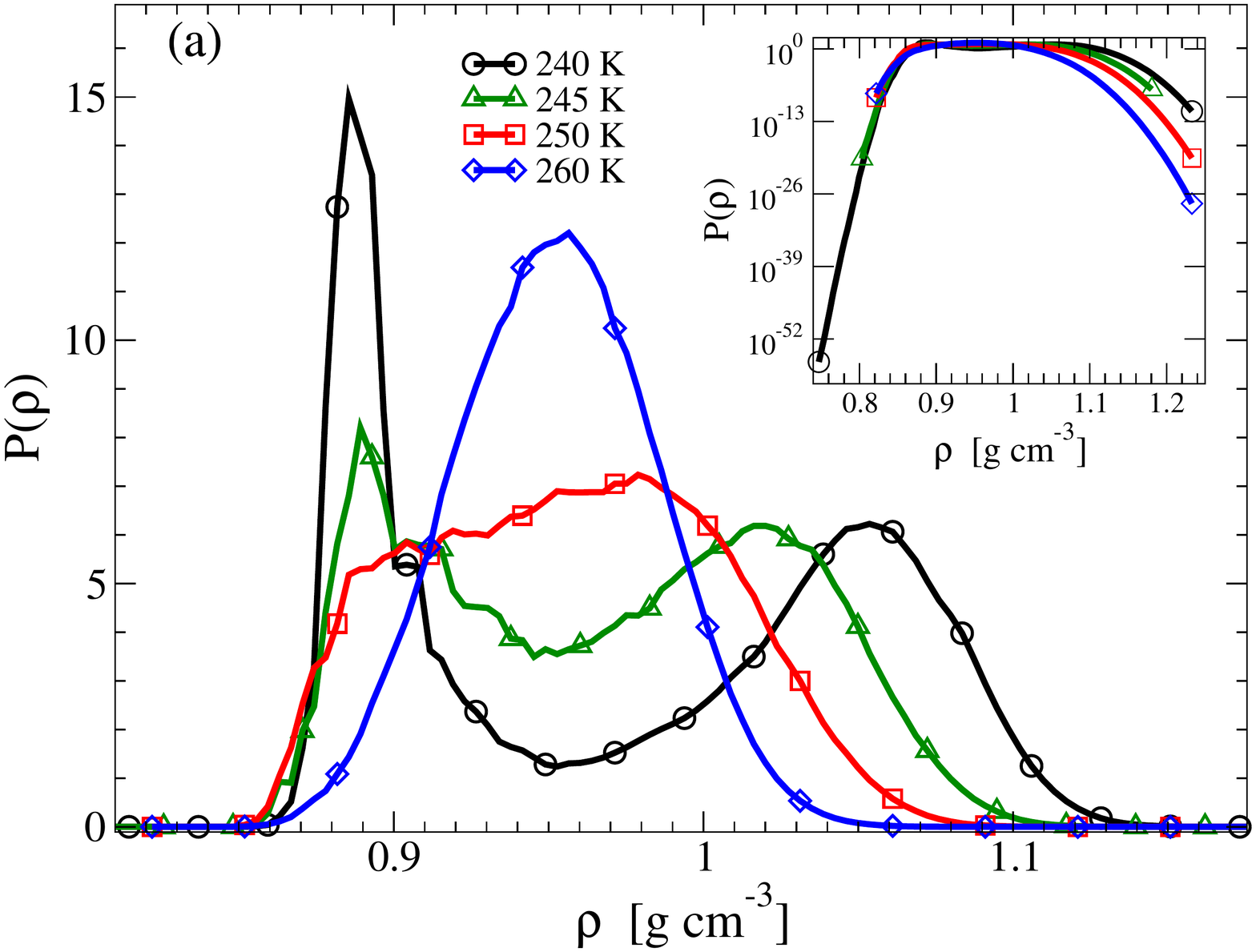}
\includegraphics[width=7.5cm]{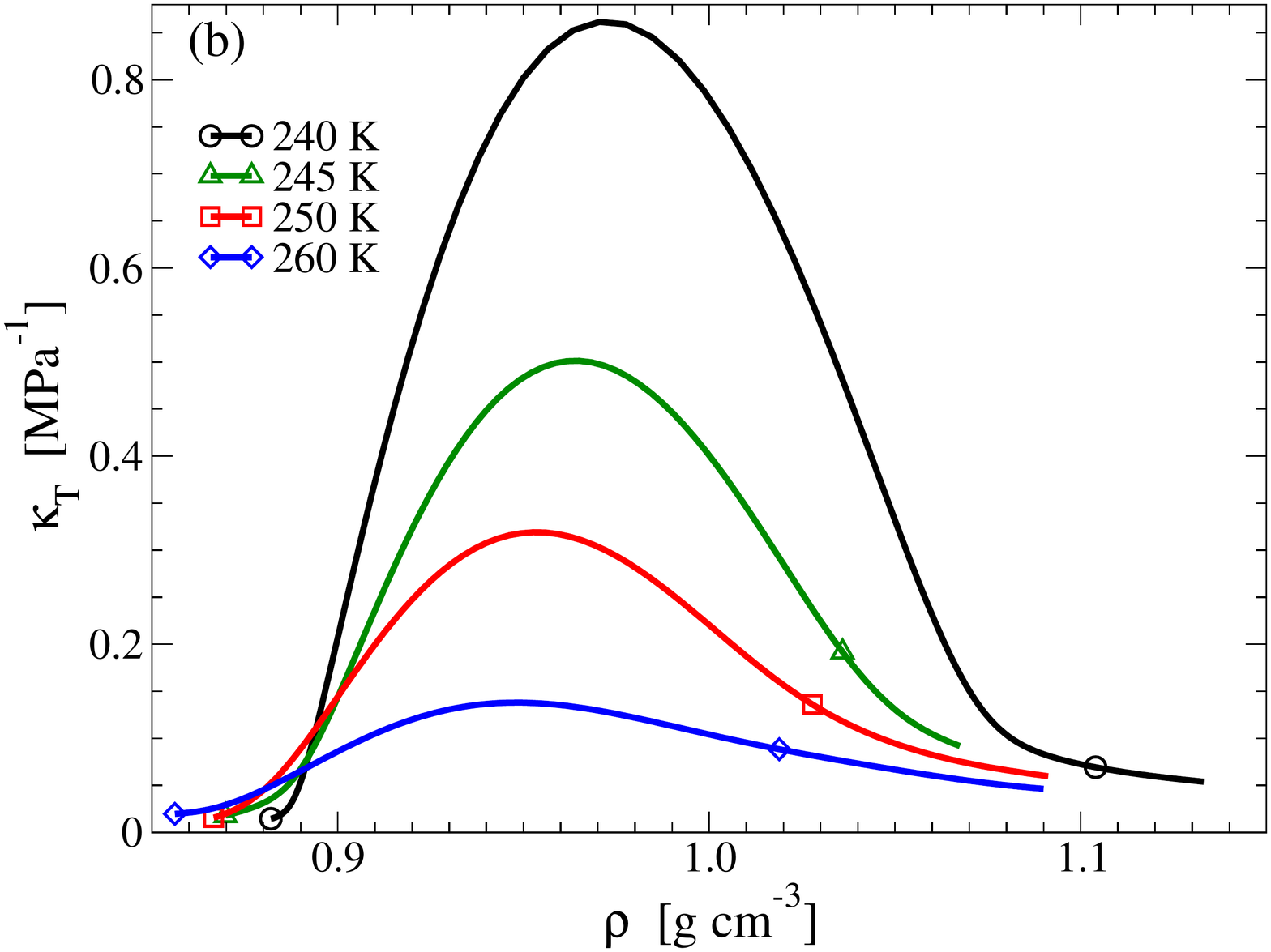}
\caption{(a) Probability distribution function of the density $P(\rho)$ obtained from SUS MC simulations
at activity values of 
$z^*= 1.145\times 10^{-4}$ for $T=240$~K,
$z^* = 1.46\times 10^{-4}$ for $T=245$~K,
$z^* = 1.85\times 10^{-4}$ for $T=250$ and  $z^* = 2.88\times 10^{-4}$  at 260~K (all in nm$^{-3}$). The corresponding values of $P$ are respectively $P=223$, 
$P=192$, $P=172$ and $P=130$ MPa. 
Below $T=250$~K, bimodality of the distribution emerges, signaling the appearance of two liquid phases with distinct densities.
(b) Isothermal compressibility as a function of density along isotherms,
obtained using Eq.~\ref{eq:compress}.
The location of the peaks  matches the data presented in Ref.~\cite{Poole:2005p2770}.
}
\label{fig:dos}
\end{figure}

\begin{figure} 
\hbox to \hsize{\epsfxsize=1\hsize\epsfbox{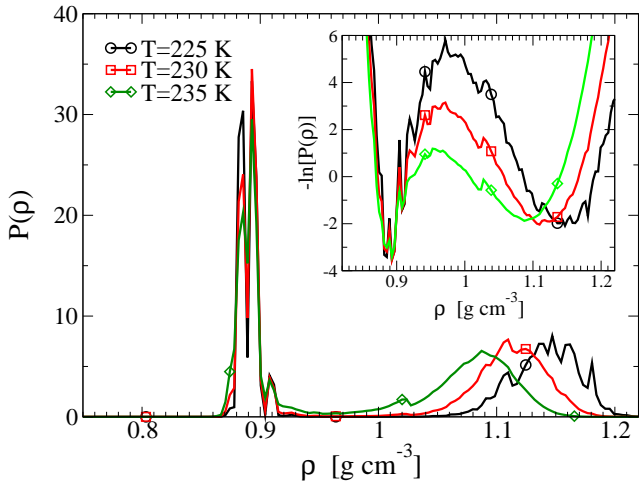}}
\caption{Density distributions $P(\rho)$ at low $T$ obtained by a reweighting of the  $T=240$ K energy-density distribution $P(\rho,E)$.  The high density peak moves to higher $\rho$ with decreasing $T$, reflecting the behavior of a simple liquid, while the location of the low density peak stays fixed because the geometric constraints for a well-formed random tetrahedral network depend sensitively on the density. Note that the numerical noise  (see e.g. the apparent minimum at 
$\rho \approx 0.89$ g/cm$^3$) is progressively amplified
reweighting at smaller and smaller $T$. }
\label{fig:coexistence}
\end{figure}

It is possible that a crystallite large enough to affect system properties, but too small to significantly affect a global measure of crystallinity 
such as $Q_6$, may be present in the system.
To test for this, we estimate the number of crystal-like particles in the system $n_{\rm crys}$.
We label particles as being crystal-like using two definitions.  For $l=3$, we define a particle as crystal-like if
it  has at least three dot products with neighboring particles satisfying $c_3^{ij} \le -0.87$.  This criterion  
has been used in nucleation studies of tetrahedral 
systems~\cite{valerianiCarbon2,ivanJCPpreprint}
and in our system the value of -0.87 is rather generous, 
in that the $c_3^{ij}$ distribution for the cubic and hexagonal crystals near $c_3^{ij}\approx -1$ falls to zero before reaching -0.87.
For $l=6$, we similarly define a particle to be crystal-like if it has three neighbors with $c_6^{ij} \ge 0.70$, a low estimate for the  
value at which the crystal distribution crosses the liquid distributions near $c_6^{ij} \approx 1$.
We plot the average fraction $n_{\rm crys}/N $ of crystal-like particles using $l=3$ and $l=6$ at $T=240$~K in Fig.~\ref{fig:cij}(c).
The largest value for $ n_{\rm crys}/N \approx 0.024$ suggests that the average crystal cluster size, in our system
of about $200-300$ molecules, is not larger than roughly six molecules,   again confirming the lack of crystallinity over the entire density range studied.  Even if we do not resort to average values, the 
largest number of crystal-like particles that we ever observe in our system, considering all of the configurations we sample, is $n_{\rm crys}=19$, found when using $l=3$.
%while the largest ever observed cluster of crystal-like particles  (defined by stating that two neighboring crystal-like particles are part of the same cluster) is thirteen. 

Our observation that crystal-like particles are rare at all densities, including in the range of the LDL phase, demonstrates that the crystal nucleation process occurs on a much longer time scale than that required for liquid-state relaxation.  Further, the fact that those small clusters of crystal-like particles that do occasionally form subsequently disappear, shows that a finite and non-trivial nucleation barrier separates the liquid phase from the crystal phase at all densities.  This demonstrates that the liquid phases simulated here are associated with free energy basins that are distinct from that of the crystal phase.

\section{Density of states}

Fig.~\ref{fig:dos}(a) shows $P(\rho)$  for all studied temperatures, at the
chemical potential for which the density fluctuations are maximal.   The data
show the onset of a bimodal distribution below $T=250$~K, consistent with the existence of a LL critical point, and suggests the occurrence of two thermodynamically distinct liquids phases associated with well-separated free energy basins.
The inset shows the same data on a semi-log scale, to highlight the ability of the SUS
method to provide an accurate estimate of $P(\rho)$ for over 50 orders of magnitude.
%For lower $T<250$~K, the two peaks are well resolved.
% and the ratio between the peak height and the intermediate valley reaches the value of 0.5, which is characteristic of Ising-like fluctuations in three dimensions at the (finite-size) critical point.  

From the energy-density probability density $P(\rho,E)$ obtained from the SUS simulations, 
all possible thermodynamic quantities
can be calculated, for all values of chemical potential (limited only by the noise quality of the data). It is interesting to determine
the behavior of the isothermal compressibility $K_T$, which in terms of the 
fluctuation in the number of particles is, 
%\begin{equation}
%k_B T \rho K_T=  \left<\Delta N^2\right>,
%\end{equation}
% {\bf could  this be}
\begin{equation}\label{eq:compress}
k_B T \rho_N K_T= \frac{\left<N^2\right> - \left<N\right>^2}{\left< N\right>},
\end{equation}
where $\rho_N$ is the number density
and $k_B$ is the Boltzmann constant. Fig.~\ref{fig:dos}(b) shows the compressibility
at several $T$.  As expected, and in agreement with previous
estimates for the $K_T$ extrema locus based on MD simulations~\cite{Poole:2005p2770}, the line of  $K_T$ maxima 
moves to smaller densities (and hence lower pressure) as $T$ increases away from the critical point.

%The $P(\rho,E)$ data at $T=240$ K can be used to estimate the densities of the coexisting phases below the critical point.  
The statistical quality of our estimate for $P(\rho,E)$ allows
us to predict $P(\rho)$ down to approximately $T=225$ K via temperature reweighting.  The
reweighted $P(\rho)$  in Fig.~\ref{fig:coexistence} shows two well-resolved peaks, with a very shallow minimum between them, indicating that the free energy barrier for the system to jump from the LDL to the HDL phase and vice versa, even in a system of only a few hundred particles, is becoming significantly larger than the thermal energy.  In the inset to Fig.~\ref{fig:coexistence}, we plot $-\ln{P(\rho)}$ to highlight this growing free energy barrier. While  the density of the HDL phase progressively increases on cooling, the density of the coexisting LDL remains essentially constant. This highlights that the thermodynamic stabilization of the LDL phase (i.e. the establishment of an equilibrium network of tetrahedrally bonded molecules) requires a very precise density to be achieved. This strong coupling between the density, local tetrahedral geometry, and free energy is at the heart of the physics of water.

\section{Discussion}

%The results reported in this article confirm the previous calculations of Liu, et al.~\cite{lpd09} and demonstrate that the critical phenomenon in ST2 water is genuine and independent of the way the long range interactions are modelled.  In addition, we provide evidence that the free energy of this model, projected onto the density, at low $T$ is characterized by two basins, both of which correspond to disordered liquid phases. Even though the LL critical point is located on the extension of the liquid free energy surface that is metastable with respect to crystal formation, the time required for homogeneous nucleation is sufficiently long in our system to allow for local equilibration in phase space. The ST2 model thus provides an unobstructed view of a LL phase transition, and its associated critical point, in a water-like system.

The results reported in this article are consistent with the previous calculations of Liu, et al.~\cite{lpd09}, and show that the evidence for the proposed critical phenomenon in ST2 water is independent of the way the long range interactions are modelled.  In addition, our results are also consistent with the possibility that the free energy of this model, projected onto the density, at low $T$ is characterized by two basins, both of which correspond to disordered liquid phases.  We also show that even though the location of the proposed LL critical point lies on the extension of the liquid free energy surface that is metastable with respect to crystal formation, the time required for homogeneous nucleation is sufficiently long in our system to allow for local equilibration in phase space.  In sum, the evidence presented here continues to point to the existence of a LL phase transition in supercooled ST2 water.

To further test for behavior consistent with a LL phase transition, a number of additional questions should be addressed.  In particular, in a finite-sized system, a bimodal density of states can occur in a non-critical system if the correlation length of the order parameter (here, the density) exceeds the system size.  A finite-size scaling analysis of the density of states, using a range of system sizes, would therefore provide an important test for the proposed LL transition in the ST2 model.  It would also be valuable to test if the free energy barrier for crystal nucleation is strongly affected by the system size, and by the constrained cubic geometry of the simulation cell, in order to more fully understand the relationship between the time scale for equilibrating the liquid and for crystal formation.
We recommend such studies for future work.

%A more precise characterization of the universality class of this LL critical point, beyond that already presented in Liu, et al.~\cite{lpd09}, can only be obtained via finite-size scaling analysis of the density of states.  This analysis can not be performed with sufficient accuracy at the present time with our available computational resources.  The confirmation that criticality can also be observed with the reaction field method (which is significantly faster than Ewald sums) gives us the hope that such a study can hopefully be attempted in the near future.

\section{acknowledgments}
We thank D. Chandler and D. Limmer for providing us with a preliminary version of Ref.~\cite{limmer3} pertaining to the interpretation of phenomena associated with a critical point in terms of crystallization.  We thank R.K. Bowles and P.G. Debenedetti for useful discussions.
FS acknowledges support from ERC-226207-PATCHYCOLLOIDS.  
IS-V thanks the Dipartimento di Fisica, {\it Sapienza}  Universit\`a di Roma, for its hospitality
and acknowledges support from ACEnet and NSERC.
PHP thanks NSERC and the CRC program.


\begin{thebibliography}{35}
\expandafter\ifx\csname natexlab\endcsname\relax\def\natexlab#1{#1}\fi
\expandafter\ifx\csname bibnamefont\endcsname\relax
  \def\bibnamefont#1{#1}\fi
\expandafter\ifx\csname bibfnamefont\endcsname\relax
  \def\bibfnamefont#1{#1}\fi
\expandafter\ifx\csname citenamefont\endcsname\relax
  \def\citenamefont#1{#1}\fi
\expandafter\ifx\csname url\endcsname\relax
  \def\url#1{\texttt{#1}}\fi
\expandafter\ifx\csname urlprefix\endcsname\relax\def\urlprefix{URL }\fi
\providecommand{\bibinfo}[2]{#2}
\providecommand{\eprint}[2][]{\url{#2}}

\bibitem[{\citenamefont{Stillinger and Rahman}(1974)}]{ST2}
\bibinfo{author}{\bibfnamefont{F.~H.} \bibnamefont{Stillinger}}
  \bibnamefont{and} \bibinfo{author}{\bibfnamefont{A.}~\bibnamefont{Rahman}},
  \bibinfo{journal}{J. Chem. Phys.} \textbf{\bibinfo{volume}{60}},
  \bibinfo{pages}{1545} (\bibinfo{year}{1974}).

\bibitem[{\citenamefont{Poole et~al.}(1992)\citenamefont{Poole, Sciortino,
  Essmann, and Stanley}}]{pses92}
\bibinfo{author}{\bibfnamefont{P.~H.} \bibnamefont{Poole}},
  \bibinfo{author}{\bibfnamefont{F.}~\bibnamefont{Sciortino}},
  \bibinfo{author}{\bibfnamefont{U.}~\bibnamefont{Essmann}}, \bibnamefont{and}
  \bibinfo{author}{\bibfnamefont{H.~E.} \bibnamefont{Stanley}},
  \bibinfo{journal}{Nature} \textbf{\bibinfo{volume}{360}},
  \bibinfo{pages}{324} (\bibinfo{year}{1992}).

\bibitem[{\citenamefont{Mishima and Stanley}(1998)}]{hes17}
\bibinfo{author}{\bibfnamefont{O.}~\bibnamefont{Mishima}} \bibnamefont{and}
  \bibinfo{author}{\bibfnamefont{H.~E.} \bibnamefont{Stanley}},
  \bibinfo{journal}{Nature} \textbf{\bibinfo{volume}{392}},
  \bibinfo{pages}{164} (\bibinfo{year}{1998}).

\bibitem[{\citenamefont{Franzese et~al.}(2001)\citenamefont{Franzese, Malescio,
  Skibinsky, Buldyrev, and Stanley}}]{hes26}
\bibinfo{author}{\bibfnamefont{G.}~\bibnamefont{Franzese}},
  \bibinfo{author}{\bibfnamefont{G.}~\bibnamefont{Malescio}},
  \bibinfo{author}{\bibfnamefont{A.}~\bibnamefont{Skibinsky}},
  \bibinfo{author}{\bibfnamefont{S.~V.} \bibnamefont{Buldyrev}},
  \bibnamefont{and} \bibinfo{author}{\bibfnamefont{H.~E.}
  \bibnamefont{Stanley}}, \bibinfo{journal}{Nature}
  \textbf{\bibinfo{volume}{409}}, \bibinfo{pages}{692} (\bibinfo{year}{2001}).

\bibitem[{\citenamefont{Abascal and Vega}(2010)}]{abascal2010}
\bibinfo{author}{\bibfnamefont{J.~L.~F.} \bibnamefont{Abascal}}
  \bibnamefont{and} \bibinfo{author}{\bibfnamefont{C.}~\bibnamefont{Vega}},
  \bibinfo{journal}{J. Chem. Phys.} \textbf{\bibinfo{volume}{133}},
  \bibinfo{pages}{234502} (\bibinfo{year}{2010}).

\bibitem[{\citenamefont{Harrington et~al.}(1997)\citenamefont{Harrington,
  Zhang, Poole, Sciortino, and Stanley}}]{Harrington:1997p4374}
\bibinfo{author}{\bibfnamefont{S.}~\bibnamefont{Harrington}},
  \bibinfo{author}{\bibfnamefont{R.}~\bibnamefont{Zhang}},
  \bibinfo{author}{\bibfnamefont{P.~H.} \bibnamefont{Poole}},
  \bibinfo{author}{\bibfnamefont{F.}~\bibnamefont{Sciortino}},
  \bibnamefont{and} \bibinfo{author}{\bibfnamefont{H.}~\bibnamefont{Stanley}},
  \bibinfo{journal}{Phys. Rev. Lett.} \textbf{\bibinfo{volume}{78}},
  \bibinfo{pages}{2409} (\bibinfo{year}{1997}).

\bibitem[{\citenamefont{Soper and Ricci}(2000)}]{soper2000}
\bibinfo{author}{\bibfnamefont{A.~K.} \bibnamefont{Soper}} \bibnamefont{and}
  \bibinfo{author}{\bibfnamefont{M.~A.} \bibnamefont{Ricci}},
  \bibinfo{journal}{Phys. Rev. Lett.} \textbf{\bibinfo{volume}{84}},
  \bibinfo{pages}{2881} (\bibinfo{year}{2000}).

\bibitem[{\citenamefont{Yamada et~al.}(2002)\citenamefont{Yamada, Mossa,
  Stanley, and Sciortino}}]{hes29}
\bibinfo{author}{\bibfnamefont{M.}~\bibnamefont{Yamada}},
  \bibinfo{author}{\bibfnamefont{S.}~\bibnamefont{Mossa}},
  \bibinfo{author}{\bibfnamefont{H.~E.} \bibnamefont{Stanley}},
  \bibnamefont{and}
  \bibinfo{author}{\bibfnamefont{F.}~\bibnamefont{Sciortino}},
  \bibinfo{journal}{Phys. Rev. Lett.} \textbf{\bibinfo{volume}{88}},
  \bibinfo{pages}{195701} (\bibinfo{year}{2002}).

\bibitem[{\citenamefont{Sciortino et~al.}(2003)\citenamefont{Sciortino,
  La~Nave, and Tartaglia}}]{lanave03}
\bibinfo{author}{\bibfnamefont{F.}~\bibnamefont{Sciortino}},
  \bibinfo{author}{\bibfnamefont{E.}~\bibnamefont{La~Nave}}, \bibnamefont{and}
  \bibinfo{author}{\bibfnamefont{P.}~\bibnamefont{Tartaglia}},
  \bibinfo{journal}{Phys. Rev. Lett.} \textbf{\bibinfo{volume}{91}},
  \bibinfo{pages}{155701} (\bibinfo{year}{2003}).

\bibitem[{\citenamefont{Liu et~al.}(2005)\citenamefont{Liu, Chen, Faraone, Yen,
  and Mou}}]{mou2005}
\bibinfo{author}{\bibfnamefont{L.}~\bibnamefont{Liu}},
  \bibinfo{author}{\bibfnamefont{S.-H.} \bibnamefont{Chen}},
  \bibinfo{author}{\bibfnamefont{A.}~\bibnamefont{Faraone}},
  \bibinfo{author}{\bibfnamefont{C.-W.} \bibnamefont{Yen}}, \bibnamefont{and}
  \bibinfo{author}{\bibfnamefont{C.-Y.} \bibnamefont{Mou}},
  \bibinfo{journal}{Phys. Rev. Lett.} \textbf{\bibinfo{volume}{95}},
  \bibinfo{pages}{117802} (\bibinfo{year}{2005}).

\bibitem[{\citenamefont{Loerting and
  Giovambattista}(2006)}]{Loerting:2006p4014}
\bibinfo{author}{\bibfnamefont{T.}~\bibnamefont{Loerting}} \bibnamefont{and}
  \bibinfo{author}{\bibfnamefont{N.}~\bibnamefont{Giovambattista}},
  \bibinfo{journal}{J. Phys.: Condens. Matter} \textbf{\bibinfo{volume}{18}},
  \bibinfo{pages}{R919} (\bibinfo{year}{2006}).

\bibitem[{\citenamefont{Seidl et~al.}(2011)\citenamefont{Seidl, Elsaesser,
  Winkel, Zifferer, Mayer, and Loerting}}]{loerting}
\bibinfo{author}{\bibfnamefont{M.}~\bibnamefont{Seidl}},
  \bibinfo{author}{\bibfnamefont{M.~S.} \bibnamefont{Elsaesser}},
  \bibinfo{author}{\bibfnamefont{K.}~\bibnamefont{Winkel}},
  \bibinfo{author}{\bibfnamefont{G.}~\bibnamefont{Zifferer}},
  \bibinfo{author}{\bibfnamefont{E.}~\bibnamefont{Mayer}}, \bibnamefont{and}
  \bibinfo{author}{\bibfnamefont{T.}~\bibnamefont{Loerting}},
  \bibinfo{journal}{Phys. Rev. B} \textbf{\bibinfo{volume}{83}},
  \bibinfo{pages}{100201} (\bibinfo{year}{2011}).

\bibitem[{\citenamefont{Mishima}(1994)}]{mishima94}
\bibinfo{author}{\bibfnamefont{O.}~\bibnamefont{Mishima}}, \bibinfo{journal}{J.
  Chem. Phys.} \textbf{\bibinfo{volume}{100}}, \bibinfo{pages}{5910}
  (\bibinfo{year}{1994}).

\bibitem[{\citenamefont{Mishima}(2000)}]{mishima2000}
\bibinfo{author}{\bibfnamefont{O.}~\bibnamefont{Mishima}},
  \bibinfo{journal}{Phys. Rev. Lett.} \textbf{\bibinfo{volume}{85}},
  \bibinfo{pages}{334} (\bibinfo{year}{2000}).

\bibitem[{\citenamefont{Giovambattista
  et~al.}(2005)\citenamefont{Giovambattista, Stanley, and
  Sciortino}}]{nicolasPRE2005}
\bibinfo{author}{\bibfnamefont{N.}~\bibnamefont{Giovambattista}},
  \bibinfo{author}{\bibfnamefont{E.~H.} \bibnamefont{Stanley}},
  \bibnamefont{and}
  \bibinfo{author}{\bibfnamefont{F.}~\bibnamefont{Sciortino}},
  \bibinfo{journal}{Phys. Rev. E} \textbf{\bibinfo{volume}{72}},
  \bibinfo{pages}{031510} (\bibinfo{year}{2005}).

\bibitem[{\citenamefont{Abascal and Vega}(2005)}]{vega}
\bibinfo{author}{\bibfnamefont{J.~L.~F.} \bibnamefont{Abascal}}
  \bibnamefont{and} \bibinfo{author}{\bibfnamefont{C.}~\bibnamefont{Vega}},
  \bibinfo{journal}{J. Chem. Phys.} \textbf{\bibinfo{volume}{123}},
  \bibinfo{pages}{234505} (\bibinfo{year}{2005}).

\bibitem[{\citenamefont{Paschek}(2005)}]{paschek2005}
\bibinfo{author}{\bibfnamefont{D.}~\bibnamefont{Paschek}},
  \bibinfo{journal}{Phys. Rev. Lett.} \textbf{\bibinfo{volume}{94}},
  \bibinfo{pages}{217802} (\bibinfo{year}{2005}).

\bibitem[{\citenamefont{Corradini et~al.}(2010)\citenamefont{Corradini, Rovere,
  and Gallo}}]{gallo2010}
\bibinfo{author}{\bibfnamefont{D.}~\bibnamefont{Corradini}},
  \bibinfo{author}{\bibfnamefont{M.}~\bibnamefont{Rovere}}, \bibnamefont{and}
  \bibinfo{author}{\bibfnamefont{P.}~\bibnamefont{Gallo}}, \bibinfo{journal}{J.
  Chem. Phys.} \textbf{\bibinfo{volume}{132}}, \bibinfo{pages}{134508}
  (\bibinfo{year}{2010}).

\bibitem[{\citenamefont{Saika-Voivod et~al.}(2001)\citenamefont{Saika-Voivod,
  Sciortino, and Poole}}]{saikapre2001}
\bibinfo{author}{\bibfnamefont{I.}~\bibnamefont{Saika-Voivod}},
  \bibinfo{author}{\bibfnamefont{F.}~\bibnamefont{Sciortino}},
  \bibnamefont{and} \bibinfo{author}{\bibfnamefont{P.~H.} \bibnamefont{Poole}},
  \bibinfo{journal}{Phys. Rev. E} \textbf{\bibinfo{volume}{63}},
  \bibinfo{pages}{011202} (\bibinfo{year}{2001}).

\bibitem[{\citenamefont{Vasisht et~al.}(2011)\citenamefont{Vasisht, Saw, and
  Sastry}}]{sastrynature}
\bibinfo{author}{\bibfnamefont{V.~V.} \bibnamefont{Vasisht}},
  \bibinfo{author}{\bibfnamefont{S.}~\bibnamefont{Saw}}, \bibnamefont{and}
  \bibinfo{author}{\bibfnamefont{S.}~\bibnamefont{Sastry}},
  \bibinfo{journal}{Nat. Phys.} \textbf{\bibinfo{volume}{7}},
  \bibinfo{pages}{549} (\bibinfo{year}{2011}).

\bibitem[{\citenamefont{Sciortino}(2011)}]{nandviews}
\bibinfo{author}{\bibfnamefont{F.}~\bibnamefont{Sciortino}},
  \bibinfo{journal}{Nat. Phys.} \textbf{\bibinfo{volume}{7}},
  \bibinfo{pages}{523} (\bibinfo{year}{2011}).

\bibitem[{\citenamefont{Liu et~al.}(2009)\citenamefont{Liu, Panagiotopoulos,
  and Debenedetti}}]{lpd09}
\bibinfo{author}{\bibfnamefont{Y.}~\bibnamefont{Liu}},
  \bibinfo{author}{\bibfnamefont{A.~Z.} \bibnamefont{Panagiotopoulos}},
  \bibnamefont{and} \bibinfo{author}{\bibfnamefont{P.~G.}
  \bibnamefont{Debenedetti}}, \bibinfo{journal}{J. Chem. Phys.}
  \textbf{\bibinfo{volume}{131}}, \bibinfo{pages}{104508}
  (\bibinfo{year}{2009}).

\bibitem[{\citenamefont{Limmer and Chandler}(2011)}]{limmer3}
\bibinfo{author}{\bibfnamefont{D.~T.} \bibnamefont{Limmer}} \bibnamefont{and}
  \bibinfo{author}{\bibfnamefont{D.}~\bibnamefont{Chandler}},
  \bibinfo{journal}{arXiv:1107.0337}  (\bibinfo{year}{2011}).

\bibitem[{\citenamefont{Virnau and M\"{u}ller}(2004)}]{sus}
\bibinfo{author}{\bibfnamefont{P.}~\bibnamefont{Virnau}} \bibnamefont{and}
  \bibinfo{author}{\bibfnamefont{M.}~\bibnamefont{M\"{u}ller}},
  \bibinfo{journal}{J. Chem. Phys.} \textbf{\bibinfo{volume}{120}},
  \bibinfo{pages}{10925} (\bibinfo{year}{2004}).

\bibitem[{\citenamefont{Poole et~al.}(2005)\citenamefont{Poole, Saika-Voivod,
  and Sciortino}}]{Poole:2005p2770}
\bibinfo{author}{\bibfnamefont{P.~H.} \bibnamefont{Poole}},
  \bibinfo{author}{\bibfnamefont{I.}~\bibnamefont{Saika-Voivod}},
  \bibnamefont{and}
  \bibinfo{author}{\bibfnamefont{F.}~\bibnamefont{Sciortino}},
  \bibinfo{journal}{J. Phys.: Condens. Matter} \textbf{\bibinfo{volume}{17}},
  \bibinfo{pages}{L431} (\bibinfo{year}{2005}).

\bibitem[{\citenamefont{Cuthbertson and Poole}(2011)}]{megan}
\bibinfo{author}{\bibfnamefont{M.~J.} \bibnamefont{Cuthbertson}}
  \bibnamefont{and} \bibinfo{author}{\bibfnamefont{P.~H.} \bibnamefont{Poole}},
  \bibinfo{journal}{Phys. Rev. Lett.} \textbf{\bibinfo{volume}{106}},
  \bibinfo{pages}{115706} (\bibinfo{year}{2011}).

\bibitem[{\citenamefont{Schulz et~al.}(2003)\citenamefont{Schulz, Binder,
  M\"uller, and Landau}}]{binder}
\bibinfo{author}{\bibfnamefont{B.~J.} \bibnamefont{Schulz}},
  \bibinfo{author}{\bibfnamefont{K.}~\bibnamefont{Binder}},
  \bibinfo{author}{\bibfnamefont{M.}~\bibnamefont{M\"uller}}, \bibnamefont{and}
  \bibinfo{author}{\bibfnamefont{D.~P.} \bibnamefont{Landau}},
  \bibinfo{journal}{Phys. Rev. E} \textbf{\bibinfo{volume}{67}},
  \bibinfo{pages}{067102} (\bibinfo{year}{2003}).

\bibitem[{\citenamefont{Ferrenberg and Swendsen}(1989)}]{reweight}
\bibinfo{author}{\bibfnamefont{A.~M.} \bibnamefont{Ferrenberg}}
  \bibnamefont{and} \bibinfo{author}{\bibfnamefont{R.~H.}
  \bibnamefont{Swendsen}}, \bibinfo{journal}{Phys. Rev. Lett.}
  \textbf{\bibinfo{volume}{63}}, \bibinfo{pages}{1195} (\bibinfo{year}{1989}).

\bibitem[{\citenamefont{De~Michele et~al.}(2006)\citenamefont{De~Michele,
  Gabrielli, Tartaglia, and Sciortino}}]{pwmnoi}
\bibinfo{author}{\bibfnamefont{C.}~\bibnamefont{De~Michele}},
  \bibinfo{author}{\bibfnamefont{S.}~\bibnamefont{Gabrielli}},
  \bibinfo{author}{\bibfnamefont{P.}~\bibnamefont{Tartaglia}},
  \bibnamefont{and}
  \bibinfo{author}{\bibfnamefont{F.}~\bibnamefont{Sciortino}},
  \bibinfo{journal}{J. Phys. Chem. B} \textbf{\bibinfo{volume}{110}},
  \bibinfo{pages}{8064} (\bibinfo{year}{2006}).

\bibitem[{\citenamefont{Poole et~al.}(2011)\citenamefont{Poole, Becker,
  Sciortino, and Starr}}]{preprint}
\bibinfo{author}{\bibfnamefont{P.~H.} \bibnamefont{Poole}},
  \bibinfo{author}{\bibfnamefont{S.~R.} \bibnamefont{Becker}},
  \bibinfo{author}{\bibfnamefont{F.}~\bibnamefont{Sciortino}},
  \bibnamefont{and} \bibinfo{author}{\bibfnamefont{F.~W.} \bibnamefont{Starr}},
  \bibinfo{journal}{dx.doi.org/10.1021/jp204889m, J. Phys. Chem. B}
  \textbf{\bibinfo{volume}{xxx}}, \bibinfo{pages}{xxx} (\bibinfo{year}{2011}).

\bibitem[{\citenamefont{Becker et~al.}(2006)\citenamefont{Becker, Poole, and
  Starr}}]{Becker:2006p15}
\bibinfo{author}{\bibfnamefont{S.~R.} \bibnamefont{Becker}},
  \bibinfo{author}{\bibfnamefont{P.~H.} \bibnamefont{Poole}}, \bibnamefont{and}
  \bibinfo{author}{\bibfnamefont{F.~W.} \bibnamefont{Starr}},
  \bibinfo{journal}{Phys. Rev. Lett.} \textbf{\bibinfo{volume}{97}},
  \bibinfo{pages}{055901} (\bibinfo{year}{2006}).

\bibitem[{\citenamefont{Steinhardt et~al.}(1983)\citenamefont{Steinhardt,
  Nelson, and Ronchetti}}]{Steinhardt}
\bibinfo{author}{\bibfnamefont{P.~J.} \bibnamefont{Steinhardt}},
  \bibinfo{author}{\bibfnamefont{D.~R.} \bibnamefont{Nelson}},
  \bibnamefont{and}
  \bibinfo{author}{\bibfnamefont{M.}~\bibnamefont{Ronchetti}},
  \bibinfo{journal}{Phys. Rev. B} \textbf{\bibinfo{volume}{28}},
  \bibinfo{pages}{784} (\bibinfo{year}{1983}).

\bibitem[{\citenamefont{Saika-Voivod et~al.}(2011)\citenamefont{Saika-Voivod,
  Romano, and Sciortino}}]{ivanJCPpreprint}
\bibinfo{author}{\bibfnamefont{I.}~\bibnamefont{Saika-Voivod}},
  \bibinfo{author}{\bibfnamefont{F.}~\bibnamefont{Romano}}, \bibnamefont{and}
  \bibinfo{author}{\bibfnamefont{F.}~\bibnamefont{Sciortino}},
  \bibinfo{journal}{J. Chem. Phys.} \textbf{\bibinfo{volume}{135}},
  \bibinfo{pages}{xxx} (\bibinfo{year}{2011}).

\bibitem[{\citenamefont{Romano et~al.}(2011)\citenamefont{Romano, Sanz, and
  Sciortino}}]{Romano}
\bibinfo{author}{\bibfnamefont{F.}~\bibnamefont{Romano}},
  \bibinfo{author}{\bibfnamefont{E.}~\bibnamefont{Sanz}}, \bibnamefont{and}
  \bibinfo{author}{\bibfnamefont{F.}~\bibnamefont{Sciortino}},
  \bibinfo{journal}{J. Chem. Phys.} \textbf{\bibinfo{volume}{134}},
  \bibinfo{pages}{174502} (\bibinfo{year}{2011}).

\bibitem[{\citenamefont{Ghiringhelli et~al.}(2008)\citenamefont{Ghiringhelli,
  Valeriani, Los, Meijer, Fasolino, and Frenkel}}]{valerianiCarbon2}
\bibinfo{author}{\bibfnamefont{L.~M.} \bibnamefont{Ghiringhelli}},
  \bibinfo{author}{\bibfnamefont{C.}~\bibnamefont{Valeriani}},
  \bibinfo{author}{\bibfnamefont{J.~H.} \bibnamefont{Los}},
  \bibinfo{author}{\bibfnamefont{E.~J.} \bibnamefont{Meijer}},
  \bibinfo{author}{\bibfnamefont{A.}~\bibnamefont{Fasolino}}, \bibnamefont{and}
  \bibinfo{author}{\bibfnamefont{D.}~\bibnamefont{Frenkel}},
  \bibinfo{journal}{Mol. Phys.} \textbf{\bibinfo{volume}{106}},
  \bibinfo{pages}{2011} (\bibinfo{year}{2008}).

\end{thebibliography}
\end{document}